\documentclass{lmcs}
\pdfoutput=1

\usepackage{lastpage}
\lmcsdoi{21}{3}{26}
\lmcsheading{}{\pageref{LastPage}}{}{}%
{Apr.~22,~2024}{Sep.~10,~2025}{}

\usepackage[utf8]{inputenc}
\usepackage[numbers]{natbib}
\usepackage{amssymb,amsmath,amsthm,amsfonts}

\usepackage{complexity}
\usepackage[boxed,linesnumbered]{algorithm2e}

\newcounter{anonymous}
\usepackage{hyperref}
\usepackage{cleveref}

\definecolor{corlinks}{RGB}{200,0,0}
\definecolor{cormenu}{RGB}{200,0,0}
\definecolor{corurl}{RGB}{200,0,0}

\hypersetup{
	colorlinks=true,
	urlcolor=corlinks,
	linkcolor=corlinks,
	menucolor=cormenu,
	citecolor=corlinks,
	pdfborder= 0 0 0
}

\newcommand{\eps}{\varepsilon}

\newcommand{\eqdef}{\triangleq}

\newcommand{\bbN}{\mathbb{N}}
\newcommand{\zo}{\{0,1\}}

\newcommand{\Log}{\mathsf{Log}}

\newcommand{\IS}{\mathsf{IS}}
\newcommand{\ISonetwo}{\mathsf{IS}^1_2}
\newcommand{\Sonetwo}{\mathsf{S}^1_2}

\newcommand{\NSIZE}{\mathsf{NSIZE}}
\newcommand{\coNSIZE}{\mathsf{coNSIZE}}

\Crefname{thm}{Theorem}{Theorems}
\Crefname{lem}{Lemma}{Lemmas}
\Crefname{cor}{Corollary}{Corollaries}
\Crefname{defi}{Definition}{Definitions}
\Crefname{clm}{Claim}{Claims}
\Crefname{prop}{Proposition}{Propositions}

\begin{document}

\title[Unprovability of Bounds in \texorpdfstring{$\mathsf{IS}^1_2$}{IS}]{On the Unprovability of Circuit Size Bounds in Intuitionistic \texorpdfstring{$\mathsf{S}^1_2$}{S12}}
\titlecomment{}

\author[L.~Chen]{Lijie Chen\lmcsorcid{0000-0002-6084-4729}}[a]
\author[J.~Li]{Jiatu Li\lmcsorcid{0000-0003-2358-3141}}[b]
\author[I.~Oliveira]{Igor C.~Oliveira\lmcsorcid{0000-0003-4048-2385}}[c]

% affiliation 1 (automatically numbered a)
\address{University of California Berkeley, Berkeley, CA, USA}	%optional
% write emails for all authors having that affiliation
\email{lijiechen@berkeley.edu}  %optional

% affiliation 2 (automatically numbered b)
\address{Massachusetts Institute of Technology, Cambridge, MA, USA}	%optional
\email{jiatuli@mit.edu}  %optional

\address{University of Warwick, Coventry, UK}
\email{igor.oliveira@warwick.ac.uk}

\begin{abstract}
 We show that there is a constant $k$ such that Buss's intuitionistic theory $\mathsf{IS}^1_2$ does not prove that SAT requires co-nondeterministic circuits of size at least $n^k$. To our knowledge, this is the first unconditional unprovability result in bounded arithmetic in the context of worst-case fixed-polynomial size circuit lower bounds. We complement this result by showing that the upper bound $\mathsf{NP} \subseteq \mathsf{coNSIZE}[n^k]$ is unprovable in $\mathsf{IS}^1_2$.

 In order to establish our main result, we obtain new unconditional lower bounds against refuters that might be of independent interest. In particular, we show that there is no efficient refuter for the lower bound $\NP \nsubseteq \mathsf{i.o.}\text{-}\mathsf{coNP}/\poly$, addressing in part a question raised by Atserias \citep{Atserias06}.
\end{abstract}

\maketitle

\section{Introduction}

\subsection{Context and motivation}

Pich and Santhanam \citep{DBLP:conf/stoc/PichS21} and subsequently Li and Oliveira  \citep{LO23} have shown that certain strong complexity lower bounds are unprovable in bounded arithmetic theory $\mathsf{PV}_1$ and its extensions, such as $\mathsf{APC}_1$ and $\mathsf{S}^1_2$. For instance, \citep{DBLP:conf/stoc/PichS21} established that $\mathsf{PV}_1$ does not prove that there is a language $L \in \mathsf{NP}$ that is average-case hard against co-nondeterministic circuits of size $2^{n^{o(1)}}$. Note that, while the unprovability results of \citep{DBLP:conf/stoc/PichS21} and \citep{LO23} are unconditional, they only apply to significantly strong complexity lower bounds. 

It would be more interesting to understand the (un)provability of circuit lower bounds that are closer to major open problems from complexity theory, such as showing that $\NP$ is not contained in $\P/\poly$. To achieve this, it is necessary to develop techniques to address the following three aspects for which these lower bound statements are stronger than $\NP \nsubseteq \P/\poly$:
\begin{enumerate}
    \item The statements consider average-case instead of worst-case lower bounds.
    \item They refer to sub-exponential size instead of super-polynomial size lower bounds.
   \item The lower bound is against 
 co-nondeterministic circuits instead of deterministic circuits.
\end{enumerate}

Aiming for unconditional results, we propose the consideration of these three challenges in the more restricted (but formally necessary) setting of \emph{intuitionistic} theories of bounded arithmetic. Intuitionistic theories distinguish themselves from classical logic systems by aligning more closely with the concept of constructive proofs. Notably, intuitionistic logic systems do not presuppose the principles of the excluded middle and double negation elimination, which give rise to key inference rules in classical logic (see, e.g., \citep{applied-proof-theory} for some background and applications).\footnote{For instance, win-win arguments from complexity theory, such as the non-constructive proof of the existence of a pseudodeterministic algorithm for generating primes from \citep{DBLP:conf/stoc/OliveiraS17} by considering if $\mathsf{PSPACE} \subset \mathsf{BPP}$ or not, are not available in intuitionistic theories.} We note that connections between complexity theory and intuitionistic logic have been widely investigated (see, e.g.,~\citep{DBLP:conf/coco/Buss86a, MR1241248, buss1990note, harnik1992provably, DBLP:journals/mlq/FerreiraM98,  buss1990model, DBLP:journals/jsyml/Avigad00, moniri2008hierarchy, ghasemloo2013note, MR3564380} and references therein). 

\subsection{Results}\label{sec:intro_results}

We show that \emph{worst-case} \emph{fixed-polynomial size} lower bounds against \emph{co-nondeterministic} circuits for a language in $\mathsf{NP}$ cannot be established in $\mathrm{IS}^1_2$ \citep{DBLP:conf/coco/Buss86a}, the intuitionistic analogue of Buss's theory $\mathsf{S}^1_2$  \citep{Buss}. This unconditional result addresses aspects (A) and (B) highlighted above, in the setting of intuitionistic bounded arithmetic.\footnote{Note that showing the unprovability of fixed-polynomial size lower bounds is stronger than showing the unprovability of super-polynomial size lower bounds.} We also remark that the lower bound we consider, which is weaker than $\NP \not\subset \coNP/\poly$, is implied by the widely believed conjecture that $\PH$ does not collapse. On the other hand, we are not aware of any standard assumption that implies the strong two-sided error average-case lower bounds for $\NP$ against subexponential-size co-nondeterministic circuits considered in previous unprovability results \cite{DBLP:conf/stoc/PichS21,LO23}.

A natural question is whether one can also unconditionally show that lower bounds against deterministic circuits (instead of co-nondeterministic circuits) are unprovable in $\mathsf{IS}^1_2$, which corresponds to Item (C) in the discussion above. This question is particularly interesting in light of the existence of non-trivial lower bounds that \emph{are} provable in $\mathsf{IS}^1_2$. For instance, it is possible to show in $\mathsf{IS}^1_2$ that the parity function on $n$ input bits requires Boolean formulas of size $\Omega(n^{3/2})$, which is a consequence of its provability in $\mathsf{S}^1_2$ \citep{DBLP:conf/innovations/CarmosinoKKOT25} and a conservation result \citep[Theorem 3.17]{DBLP:journals/jsyml/Avigad00}. As we explain in more detail in \Cref{sec:lb_det}, for some formalizations of lower bound statements against deterministic circuits, unprovability in $\mathsf{IS}^1_2$ yields unprovability in the stronger theory $\mathsf{S}^1_2$, an observation that first appeared in Ghasemloo and Pich \citep{ghasemloo2013note}. We believe that this connection  further motivates the investigation of the unprovability of complexity lower bounds in intuitionistic theories.

Next, we provide more details about our results and formalizations.

\paragraph*{Formalizations.} Fix a polynomial-time nondeterministic machine $M$. Given $n_0 \in \mathbb{N}$  and a size-bound $s \colon \mathbb{N} \to \mathbb{N}$ that is sufficiently time-constructive and satisfies $n \leq s(n) \leq 2^n$, we consider a sentence $\mathsf{LB}^{\mathsf{exp}}(M,s,n_0)$ stating that:\footnote{Note that $n\in\Log\Log$ essentially means that bounded quantifiers refer to objects of length $2^{O(n)}$; similarly, $n\in\Log$ means that bounded quantifiers refer to objects of length $\poly(n)$. For a formal definition, see \Cref{sec:prelim_intuitionistic}.}
$$
\forall n \in \mathsf{LogLog}~\text{with}~n \geq n_0~\forall D \in \mathsf{coNSIZE}[s(n)]~\exists x \in \{0,1\}^n~\mathsf{Error}_M(D,n,x)\;,
$$
where $\mathsf{Error}_M(D,n,x)$ denotes the formula
$$
(\exists y,z \in \{0,1\}^{s(n)}\,M(x,y) = 1 \wedge D(x,z) = 0) \vee (\forall y,z \in \{0,1\}^{s(n)} ~M(x,y) = 0 \wedge D(x,z) = 1)\;.
$$
This sentence expresses that, for every $n \geq n_0$, no co-nondeterministic circuit of size at most $s(n)$ decides $L(M)$ on inputs of length $n$. The notation $\mathsf{LB}^{\mathsf{exp}}$ emphasizes that a candidate proof of this sentence is able to manipulate concepts of exponential size $2^{O(n)}$, since $n \in \mathsf{LogLog}$. We refer to \Cref{sec:prelim_intuitionistic} for more details on how to formalize this sentence in the setting of $\mathsf{IS}^1_2$.

If $s(n) = n^k$ for some rational number $k \geq 1$, we can consider a sentence $\mathsf{LB}^{\mathsf{poly}}(M,s,n_0)$ stating that:
$$
\forall n \in \mathsf{Log}~\text{with}~n \geq n_0~\forall D \in \mathsf{coNSIZE}[s(n)]~\exists x \in \{0,1\}^n~\mathsf{Error}_M(D,n,x)\;.
$$
The key difference here is that when $s(n) = n^k$ it is sufficient to assume $n \in \mathsf{Log}$ to obtain a natural  formalization of complexity lower bounds. Correspondingly, the notation $\mathsf{LB}^{\mathsf{poly}}$ emphasizes that a candidate proof of this sentence can manipulate concepts of size polynomial in $n$.

\begin{thm} \label{thm:intro_unprovIS12} The following results hold:
\begin{enumerate}[label=(\roman*)]
    \item \emph{[Exponential Regime]} Let $\delta > 0$ be a rational number, $n_0 \in \mathbb{N}$, and $M$ be a polynomial-time nondeterministic machine. Then
    $
  \mathsf{IS}^1_2 \nvdash   \mathsf{LB}^{\mathsf{exp}}(M,2^{n^{\delta}},n_0)\;.
$
    \item \emph{[Polynomial Regime]} Let $n_0 \in \mathbb{N}$ and $M$ be a polynomial-time nondeterministic machine. Then there is an integer $k  \geq 1$ such that
    $
  \mathsf{IS}^1_2 \nvdash   \mathsf{LB}^{\mathsf{poly}}(M,n^k,n_0)\;.
  $
\end{enumerate}
\end{thm}

The proof of \Cref{thm:intro_unprovIS12} combines two main steps. First, we employ a strong witnessing result \citep{DBLP:conf/coco/Buss86a, MR1241248} for the intuitionistic theory  $\mathsf{IS}^1_2$ to show that the provability of a circuit lower bound yields a computationally bounded \emph{refuter}. A refuter for a lower bound of the form $L \notin \mathcal{C}$, where $\mathcal{C}$ is a complexity class, is an algorithm $R(1^n,E)$ that, given an input length $n$ and a device $E$ from $\mathcal{C}$, outputs an $n$-bit string $x$ such that $E(x) \neq L(x)$. We then establish unconditionally that such a refuter does not exist. Consequently, the lower bound sentence cannot be proved in the theory  $\mathsf{IS}^1_2$.

Elaborating on our approach, we now present some new consequences for refuters that might be of independent interest. We use $\mathsf{SAT}_n$ to denote the decision version of the satisfiability problem for De Morgan Boolean formulas represented by $n$-bit strings (our results are robust to encoding details). As in the discussion above, we say that a refuter $R$ for a language $L$ succeeds against a class $\mathcal{C}$ of devices  on input length $n$ if, for every $E \in \mathcal{C}$,  $R(1^n,E)$ outputs $x \in \{0,1\}^n$ such that $E(x) \neq L(x)$. Observe that the existence of such a refuter implies that $L \notin \mathcal{C}$.

\begin{thm} \label{thm:intro_refuters}
The following results hold:
\begin{enumerate}[label=(\roman*)]
     \item Let $k$ and $c$ be rational numbers such that $1 < k < c < k^2$. Then there is no non-uniform refuter $R(1^n,E_n)$ for $\mathsf{SAT}_n$ against $E_n \in \mathsf{coNSIZE}[n^k]$ that has circuit size $\leq n^c$ and succeeds on every large enough input length $n$.
     \item There is an integer $k \geq 1$ such that there is no non-uniform refuter $R(1^n,E_n)$ for $\mathsf{SAT}_n$ against $E_n \in \mathsf{coNSIZE}[n^k]$ of polynomial circuit size which succeeds on every large enough input length.
\end{enumerate}
\end{thm}

These two items are incomparable. The first item of \Cref{thm:intro_refuters} is an impossibility result for refuting even very weak lower bounds ($n^k$ gates for any fixed $k > 1$), but only addresses refuters of size smaller than $n^{k^2}$. On the other hand, the second item holds against any  polynomial-size refuter, but does not provide an explicit constant $k$ for the size bound.

Atserias \citep{Atserias06} raised the
following related questions in a work showing that there is a randomized refuter for the lower bound $\NP \nsubseteq \P/\poly$ (assuming $\NP \nsubseteq \P/\poly$):\footnote{As usual, for a set $\mathcal{C}$ of languages, we let $\mathsf{i.o.}\text{-}\mathcal{C}$ denote the set of languages $L'$ for which there is some $L \in \mathcal{C}$ such that $L'$ and $L$ agree on infinitely many input lengths.}
\begin{itemize}
    \item Is there a refuter for $\NP \nsubseteq \mathsf{i.o.}\text{-}\P/\poly$?
    \item Is there a refuter for $\NP \nsubseteq \mathsf{coNP}/\poly$?
\end{itemize}
Thus, \Cref{thm:intro_refuters} Item (\emph{ii}) provides an answer to a combination of these questions: there is no refuter for $\NP \not\subset \mathsf{i.o.}\text{-}\mathsf{coNP}/\poly$.

At a high level, the proof of \Cref{thm:intro_refuters} proceeds by contradiction. From the existence of such a refuter, we obtain a worst-case \emph{upper bound} on the complexity of $\mathsf{SAT}_n$. This step requires an extension of a technique from previous papers \citep{DBLP:journals/jml/Krajicek11, DBLP:journals/apal/Pich15} to the setting of \emph{worst-case complexity} and to the \emph{polynomial circuit size} regime. To achieve this, we employ a new \emph{bootstrapping argument} (see \Cref{sec:refuters}) that invokes the refuter over different input lengths and aggregates the information obtained from it. Finally, the circuit size upper bound extracted from the refuter contradicts the original assumption that a refuter exists, since it implicitly assumes a  corresponding circuit lower bound. We note that the second item of \Cref{thm:intro_refuters} requires an extra non-constructive ingredient, and as a result, the proof does not produce an explicit constant $k$.

Finally, we complement \Cref{thm:intro_unprovIS12} by establishing the unprovability in $\mathsf{IS}^1_2$ of the upper bound $\NP\subseteq \mathsf{coNSIZE}[n^k]$, for any fixed $k\in\bbN$. The formalization of this upper bound statement is presented in \Cref{ref:unprov_upper_bounds}.

\begin{thm}[Informal, see \Cref{thm:unprovIS12upper}]\label{thm:intro unprov upper bound}
    For any constant $k\in\bbN$, $\IS^1_2\nvdash``\NP\subseteq\mathsf{coNSIZE}[n^k]"$.
\end{thm}

The proof of \Cref{thm:intro unprov upper bound} relies on an adaptation of an approach from \citep{KO17}, which reduces the \emph{unprovability of non-uniform circuit upper bounds} to establishing certain \emph{uniform circuit lower bounds} (in the standard sense of complexity theory). In our result, the required uniform circuit lower bound is achieved by a modification of an argument from \citep{DBLP:journals/cc/SanthanamW14}.

\Cref{thm:intro unprov upper bound} contributes to an active line of research on the unprovability of circuit upper bounds \citep{DBLP:journals/jsyml/CookK07, KO17, DBLP:journals/aml/BydzovskyM20, BKO20, CKKO21, DBLP:conf/stoc/AtseriasBM23}. In contrast to other unconditional results from  these papers, which hold even for stronger theories, \Cref{thm:intro unprov upper bound} establishes the unprovability of \emph{co-nondeterministic} circuit upper bounds for $\mathsf{NP}$.\footnote{Note that establishing the unprovability of co-nondeterministic circuit size upper bounds is stronger than establishing the unprovability of deterministic circuit size upper bounds.} 

Altogether, \Cref{thm:intro_unprovIS12} Item (\emph{ii}) and \Cref{thm:intro unprov upper bound} bring us closer to an unconditional \emph{independence} result for $\mathsf{IS}^1_2$ with respect to worst-case fixed-polynomial bounds in co-nondeterministic circuit complexity.\footnote{Formally, our results show that for each $L \in \mathsf{NP}$ there is $k$ such that the lower bound $L \notin \mathsf{coNSIZE}[n^k]$ is unprovable, and that for each $k$ there is $L \in \mathsf{NP}$ such that the upper bound $L \in \mathsf{coNSIZE}[n^k]$ is unprovable. While our unprovability results are robust to the use of different machines of the same time complexity to represent $L$ in a sentence, we note that they do not give a fixed pair $(L',k')$ for which both lower bounds and upper bounds are unprovable.}

Finally, we note that both \Cref{thm:intro_unprovIS12} and \Cref{thm:intro unprov upper bound} can be extended to ``semi-classical'' formalizations of the statements, where the sub-formula inside the outermost existential quantifier can be replaced by any classically equivalent formula. In particular, the inner part of the lower bound sentence can be replaced with its double negation translation. Details of this extension are discussed in \Cref{sec:unprov_upper_bound_extension}. 

\subsection{Related work} 

Ghasemloo and Pich \citep{ghasemloo2013note} studied connections between natural proofs \citep{DBLP:journals/jcss/RazborovR97} and intuitionistic logic. In a bit more detail, the theory of natural proofs can be used to establish the \emph{conditional} unprovability of circuit lower bounds in classical theories admitting certain interpolation theorems (see \citep{razborov1995unprovability, krajivcek1997interpolation}), and \citep{ghasemloo2013note} investigates what (conditional) consequences this can have for the provability of lower bounds in intuitionistic theories.

\Cref{thm:intro_unprovIS12} should be contrasted with a result from Cook and Urquhart \citep[Theorem 10.16]{MR1241248} establishing the unprovability of super-polynomial lower bounds for the extended Frege propositional proof system in the related intuitionistic theory  $\mathsf{IPV}^{\omega}$ (see also \citep[Corollary 4]{krajivcek1990propositional} and \citep[Section 6.2]{buss1990model}). We note that the two results consider different lower bound questions, formalizations, and choice of parameters. Moreover,  their proofs rely on  completely different approaches. For instance, in terms of the formalization, the sentences $\mathsf{LB}^{\mathsf{exp}}$ and $\mathsf{LB}^{\mathsf{poly}}$ do not state that the input $x$ under $\mathsf{Error}_M(x)$ is a tautology, which appears to be crucial in the results from \citep{MR1241248, krajivcek1990propositional}. Furthermore, to our knowledge, it is not known how to extend their results to the fixed-polynomial size regime, as in \Cref{thm:intro_unprovIS12} Item (\emph{ii}).

It is perhaps interesting to compare \Cref{thm:intro_refuters} Item (\emph{ii}) with the positive results of Gutfreund, Shaltiel, and Ta-Shma \citep{GutfreundST07} about the existence of refuters (see also \citep{CJSW21} and references therein). For instance, \citep[Lemma 4.1]{GutfreundST07} roughly states that, if $\mathsf{NP} \neq \mathsf{RP}$, then it is possible to produce counter-examples to the correctness of a randomized input machine of complexity $n^k$ in time of order $n^{k^2}$. While the parameters of our lower bound and of their upper bound nearly match, we note that their results and \Cref{thm:intro_refuters} Item (\emph{ii}) refer to different complexity lower bounds and consider slightly different refuter guarantees.

Finally, we mention that the bootstrapping procedure employed in the proof of \Cref{thm:intro_refuters} is somewhat similar to an argument that appears in \citep{DBLP:conf/innovations/BogdanovTW10} in the context of refuters for SAT solvers.

\section{Preliminaries}\label{sec:prelim}

\newcommand{\qref}[1]{(\ref{#1})}
\newcommand{\io}{\mathsf{i.o.}\text{-}}

\subsection{Intuitionistic bounded arithmetic}\label{sec:prelim_intuitionistic}

 We consider the intuitionistic theory of bounded arithmetic $\mathsf{IS}^1_2$ introduced in \citep{DBLP:conf/coco/Buss86a} (see also \citep{MR1241248, buss1990note} for equivalent definitions). Our unprovability results are quite robust, as they rely on the consequences of a witnessing theorem for $\mathsf{IS}^1_2$ and do not crucially depend on details of the formalization. Nevertheless, for concreteness and in order to complement the discussion in \Cref{sec:lb_det}, below we provide more details about the theory. The exposition assumes basic familiarity with bounded arithmetic (see, e.g., \citep{buss-survey} for the necessary background).

 Informally, $\ISonetwo$ can be defined as theory $\Sonetwo$ but with intuitionistic predicate logic and polynomial induction restricted to $\Sigma^{b+}_1$-formulas, i.e., $\Sigma^{b}_1$-formulas that do not contain implications and negations. Next, we review some  details for a reader that might not be familiar with this terminology. Since we will not make use of sequent calculus, we follow the Hilbert-style equivalent presentation from \citep{MR1241248}.

 As in the case of $\mathsf{S}^1_2$, we take the language of $\mathsf{IS}^1_2$ to consist of non-logical symbols $0$, $S$, $+$, $\times$, $\#$, $\lfloor \frac{1}{2} x \rfloor$, $|x|$, and $\leq$ with their usual  interpretations over the intended standard model $\mathbb{N}$. The logical symbols are $\wedge$, $\vee$, $\rightarrow$, $\forall$, $\exists$, and $=$. The connectives $\neg$ and $\leftrightarrow$ can be introduced with appropriate abbreviations using $\rightarrow$ \citep[Page 109]{MR1241248}. In terms of non-logical axioms,  $\mathsf{IS}^1_2$ consists of 21 basic axioms (see \citep[Page 111]{MR1241248}) and the axiom scheme $\Sigma^{b+}_1$-PIND (discussed below). The standard logical axiom schemes and rules of inference governing intuitionistic predicate logic are listed  in \citep[Page 110]{MR1241248}.

For a term $t$ not containing a variable $x$, we can define \emph{bounded quantifiers} $(\exists x \leq t)\, \varphi$ and $(\forall x \leq t)\, \varphi$ via the abbreviations $\exists x\,(x \leq t \wedge \varphi)$ and $\forall x\,(St \leq x \vee \varphi)$, respectively. \emph{Sharply bounded quantifiers} are bounded quantifiers of the form $(\exists x \leq |t|)$ and $(\forall x \leq |t|)$. We recall that $\Pi^b_0 = \Sigma^b_0$ is the set of formulas whose quantifiers are all sharply bounded. Similarly, formulas containing only bounded quantifiers can be classified into  hierarchies $\Sigma^b_i$ and $\Pi^b_i$ by counting alternations of bounded quantifiers while ignoring sharply bounded quantifiers (cf.~\citep[Page 111]{MR1241248}). A formula is \emph{positive} if it contains no occurrence of the symbol $\rightarrow$ (recall that negations are introduced via abbreviation and do not need to be explicitly discussed). A formula is $\Sigma^{b+}_1$ if it is both $\Sigma^b_1$ and positive.

%\jiatu{Adding the definition of $\Log$:}

We define $\forall n\in\Log$ (resp.~$\exists n\in\Log$) as the abbreviation of $\forall N~\forall n=|N|$ (resp.~$\exists N~\exists n=|N|$). Similarly, we define $\forall n\in\Log\Log$ (resp.~$\exists n\in\Log\Log$) as the abbreviation of $\forall N~\forall n=||N||$ (resp.~$\exists N~\exists n=||N||$). 

Finally, $\mathsf{IS}^1_2$ admits the aforementioned $\Sigma^{b+}_1$-PIND axiom scheme, consisting of formulas (possibly with additional parameters) of the form
$
\varphi(0) \wedge \forall x \,(\varphi(\lfloor \frac{1}{2} x \rfloor) \rightarrow \varphi(x))\rightarrow \forall y\,\varphi(y)\;,
$
where $\varphi$ is a $\Sigma^{b+}_1$-formula.

\paragraph*{Formalization of complexity lower bounds.} We now discuss the formalization of the sentences $\mathsf{LB}^{\mathsf{exp}}(M,s,n_0)$ and $\mathsf{LB}^{\mathsf{poly}}(M,s,n_0)$ informally introduced in \Cref{sec:intro_results}. Recall that $M$ is a non-deterministic polynomial-time machine, $s \colon \mathbb{N} \to \mathbb{N}$, and $n_0 \in \mathbb{N}$. The fixed machine $M$ and the constant $n_0$ can be explicitly encoded using a term of $\mathsf{IS}^1_2$ built from the constant symbol $0$ and from the other function symbols. Note that, in order to formalize the lower bound statements, it is sufficient to be able to define any polynomial-time function in the language of $\mathsf{IS}^1_2$. Indeed, this allows us to employ appropriate formulas to specify the output of $M$ on a given input, check if an object $D$  encodes a circuit of a given size, evaluate a given circuit on an input pair $(x,y)$, decode an $n$-bit string from an object $x$, etc. 

It turns out that, as established in \citep[Corollary 2.7]{MR1241248}, every function $f$ computable in polynomial time is $\Sigma^{b+}_1$-definable in $\mathsf{IS}^1_2$. This means that there is a $\Sigma^{b+}_1$-formula $\phi(x,y)$ such that $\mathbb{N} \models \forall x\,\phi(x,f(x))$ and $\mathsf{IS}^1_2$ proves that $\forall x\, \exists !\, y\,\phi(x,y)$ (see \citep[Pages 114-115]{MR1241248}). Given this result, it is not hard to fully specify the sentences $\mathsf{LB}^{\mathsf{exp}}(M,s,n_0)$ and $\mathsf{LB}^{\mathsf{poly}}(M,s,n_0)$ using a  formula for each relevant polynomial-time function. Since the details of how this can be done appear on previous works on the provability of lower bounds in bounded arithmetic (see, e.g., \citep{DBLP:journals/apal/Pich15}), below we comment only on the part of the formalization that affects the running time of computations obtained from proofs in $\mathsf{IS}^1_2$.

For instance, the sentence $\mathsf{LB}^{\mathsf{exp}}(M,s,n_0)$ can be expressed as
$$
\forall N\, \forall n \, \forall D\, \exists x\, \Big (n = ||N|| \wedge n \geq n_0  \wedge |x| = n \wedge \mathsf{Circuit}_{s}(D,n) \rightarrow \mathsf{Error}_M(D,n,x)\Big )\;,
$$
where $\mathsf{Circuit}_s$ is a $\Sigma^{b+}_1$-formula that checks if $D$ is the description of a co-nondeterministic circuit of size at most $s(n)$, and $\mathsf{Error}_M$ is built as a formula that checks if $M(x) \neq D(x)$. Note that if $x$ is an $n$-bit number and $D$ encodes a circuit of size at most $2^n$, a polynomial-time function $f$ over inputs $N$, $n$, and $D$ computes in time $\mathsf{poly}(|N|,|D|,|n|) = 2^{O(n)}$. On the other hand, when $s(n) \leq n^k$ for some $k$, we can take $n = |N|$ in the formalization of $\mathsf{LB}^{\mathsf{poly}}(M,s,n_0)$, which yields an algorithm from a proof   in $\mathsf{IS}^1_2$ whose running time is $\mathsf{poly}(n)$ instead of $\mathsf{poly}(2^n)$.

We stress that our unprovability results do not depend on the specific formulas employed to formalize the lower bound sentence. For more detailed discussion, see \Cref{sec:unprov_upper_bound_extension}.

\subsection{Complexity theory}

We assume basic familiarity with complexity theory, e.g., complexity classes $\P,\NP$ and $\coNP$ (see \cite{AB09}). We use $\NSIZE[s(n)]$ (resp.~$\coNSIZE[s(n)]$) to denote the set of languages decidable by families of nondeterministic (resp.~co-nondeterministic) circuits of size $s(n)$. For a set $\mathcal{C}$ of languages, we let $\mathsf{i.o.}\text{-}\mathcal{C}$ denote the set of languages $L'$ for which there is some $L \in \mathcal{C}$ such that $L'$ and $L$ agree on infinitely many input lengths.

\paragraph{Uniform circuits.} Let $\mathcal{C}$ be a non-uniform complexity class, e.g., $\mathcal{C}=\SIZE[\poly(n)]$ or $\mathcal{C}=\NSIZE[\poly(n)]$. We say that $L\in\P\text{-}\mathsf{uniform}~\mathcal{C}$ if $L$ is decidable by a family of $\mathcal{C}$-circuits $\{C_n:\zo^n\to\zo\}_{n\in\mathbb{N}}$ such that there is a polynomial-time Turing machine $M$ such that $M(1^n)$ outputs the description of $C_n$. 

\paragraph{Refuters.} Let $L$ be a language, $R$ be an algorithm, and $\mathcal{C}$ be a complexity class. We often abuse notation and also view $\mathcal{C}$ as a class of computational devices. We say that  $R$ is a refuter for the lower bound $L \notin \mathcal{C}$ (or a refuter for $L$ against $\mathcal{C}$) on input length $n$ if, for every $E \in \mathcal{C}$,  $R(1^n,E)$ outputs $x \in \{0,1\}^n$ such that $E(x) \neq L(x)$. We extend this definition in the natural way when $R$ represents a non-uniform family of circuits. In this case, we might omit the input $1^n$ and simply write $R_n$.

\section{Unprovability of Lower Bounds in \texorpdfstring{$\mathsf{IS}^1_2$}{IS12}}

In this section, we prove new results about refuters and employ them to establish the unprovability of lower bounds against co-nondeterministic circuits in theory $\mathsf{IS}^1_2$.

\subsection{Unconditional lower bounds for refuters}\label{sec:refuters}

In this section, we prove each item of \Cref{thm:intro_refuters} and establish some related results needed in the proof of \Cref{thm:intro_unprovIS12}. First, we prove the following  lemma, which shows that \emph{worst-case} upper bounds can be extracted from refuters, even in situations where the refuter runs in exponential time. The lemma can be used to show an unconditional lower bound against refuters and will be useful in the proof of \Cref{thm:intro_unprovIS12} Item (\emph{i}).

\begin{lem} \label{thm:UB_from_refuter} Let $L \in \mathsf{NP}$, and let $\delta > 0$. Suppose that there is an algorithm $R(1^n,D)$ such that, for every co-nondeterministic circuit $D$ on $n$ input variables and of size at most $2^{n^{\delta}}$, $R(1^n,D)$ runs in time $2^{O(n)}$ and outputs a string $w \in \{0,1\}^n$ such that $D(w) \neq L(w)$. Then, for every language $L' \in \mathsf{NP}$ and for every constant $\varepsilon > 0$, we have $L' \in \mathsf{DTIME}[2^{n^{\varepsilon}}]$.
\end{lem}

\begin{proof} Suppose that $L \in \mathsf{NTIME}[n^d]$ for some $d \in \mathbb{N}$. Let $M'$ be a nondeterministic machine that decides $L'$ and runs in time at most $n^{c'}$, where $c' \in \mathbb{N}$. Let $\varepsilon > 0$ be an arbitrary constant. Finally, let $\gamma = \gamma(d, \varepsilon) > 0$ be a small enough constant to be defined later. We argue that the following deterministic algorithm $B^\gamma(x)$ decides $L'$ in time  $O(2^{n^{\varepsilon}})$:
\begin{enumerate}
    \item  Let $x \in \{0,1\}^n$ be the input string.
    \item $B^\gamma$ computes the description of a co-nondeterministic circuit $E'$  of size at most $n^{2c'}$ that decides the complement of $L'$. In other words, $E'(u) = 1 - L'(u)$ for every string $u \in \{0,1\}^n$.
    \item $B^\gamma$ produces the code of a co-nondeterministic circuit $D_x(y)$, where $y \in \{0,1\}^{n^\gamma}$, such that $D_x(y)$  ignores its input $y$ and computes according to $E'(x)$.

    (In other words, while the length of the main input string $y$ of $D_x(y)$ is smaller than the length of the main input string $u$ of $E'(u)$, they share the same non-deterministic input string, and $E'$ sets $u$ to be the fixed string $x$.)  
    \item $B^\gamma$ computes $w = R(1^{n^\gamma}, D_x) \in \{0,1\}^{n^\gamma}$.
    \item Finally, $B^\gamma$ determines the bit $b = L(w)$ by a brute force computation, then sets $b$ as its output bit.
\end{enumerate}

First, we argue that $B^\gamma$ decides $L'$. Since $D_x$ is a co-nondeterministic circuit over $m = n^{\gamma}$ input strings and of size at most $n^{2c'} = m^{2c'/\gamma} \leq 2^{m^{\delta}}$ (for a large enough $m$ and assuming that $\gamma$ is constant),   $R(1^{n^\gamma}, D_x)$ outputs a string $w \in \{0,1\}^{n^\gamma}$ such that $L(w) = 1 - D_x(w)$. Consequently,
$$
b = L(w) = 1 - D_x(w) = 1 - E'(x) = 1 - (1 - L'(x))  = L'(x)\;,
$$
i.e., the output bit of $B(x)$ is correct.

Next, we argue that $B$ runs in time at most $O(2^{n^{\varepsilon}})$. Clearly, Steps 1--3 all run in $\mathsf{poly}(n)$ time. Moreover,  Step 4 runs in time $2^{O(n^{\gamma})}$ under the assumption on the running time of $R(1^{n^\gamma}, D_x)$. This is at most $2^{n^\varepsilon}$ if we set $\gamma \leq \varepsilon/2$. Finally, since $L \in \mathsf{NTIME}[n^d]$, the brute force computation in Step 5 can be performed in deterministic time $2^{O(\ell^{d})}$ over an input of length $\ell$. Since $\ell = n^{\gamma} = |w|$ in our case, if $\gamma \leq \varepsilon/2d$ we get that Step $5$ runs in time at most $2^{n^{\varepsilon}}$. Overall, if we set $\gamma \eqdef \varepsilon/2d$, it follows that $B^\gamma$ runs in time at most $O(2^{n^\varepsilon})$. This completes the proof that $L' \in \mathsf{DTIME}[2^{n^\varepsilon}]$.\footnote{We observe that, in the proof of \Cref{thm:UB_from_refuter}, it is enough for the refuter $R$ to work on co-nondeterministic circuits of size $n^{\alpha(n)}$, where $\alpha(n) \to \infty$. However, the formulation above will be sufficient for our purposes.}
\end{proof}

In the proof of the next result, we employ a more sophisticated \emph{bootstrapping argument} consisting of iterated  applications of the refuter over different input lengths. Recall that we use $\mathsf{SAT}_n$ to denote the satisfiability problem for De Morgan Boolean formulas represented by $n$-bit strings.

\begin{thm}[Restatement of \Cref{thm:intro_refuters} Item (\emph{i})]\label{thm:norefuter}
Let $k$ and $c$ be rational numbers such that $1 < k < c < k^2$. Then there is no non-uniform refuter $R(1^n,E_n)$ for $\mathsf{SAT}_n$ against $E_n \in \mathsf{coNSIZE}[n^k]$ that has circuit size $\leq n^c$ and succeeds on every large enough input length $n$.
\end{thm}

\begin{proof}
In order to simplify some calculations, we prove the result for a fixed but arbitrary language $L \in \mathsf{NTIME}[n]$. Since $\mathsf{poly}(\log n)$  overheads in our estimates do not affect the result, it is easy to check that the same proof works for a language in $\mathsf{NTIME}[n \cdot \mathsf{poly}(\log n)]$, such as $\mathsf{SAT} = \{\mathsf{SAT}_n\}$. Similarly, we will tacitly assume that machines of complexity $t$ can be simulated by circuit of size $t$ (instead of $O(t \cdot \log t)$). Our construction and upper bounds can be easily adjusted to account for these small overheads, since for constants $a < b$, $n^a \cdot \mathsf{poly}(\log n) < n^b$ for every large enough $n$. 

Let $1 < k < c < k^2$, and let $M$ be a linear time nondeterministic machine that decides $L$. Now consider an arbitrary $n_0 \in \mathbb{N}$, and suppose towards a contradiction that $R = \{R_\ell\}_{\ell \geq n_0}$ is a sequence of nonuniform circuits $R_\ell$ of size $\leq \ell^c$ such that, given a circuit $E_\ell \in \mathsf{coNSIZE}[\ell^k]$ over $\ell$ input bits, $R_\ell(E_\ell)$ outputs a string $x_\ell$ such that $M(x_\ell) \neq E_\ell(x_\ell)$. We will prove the following claim.

\begin{clm}\label{claim:upper_from_refuter}
    There is an input length $\ell_0 \geq n_0$ and a deterministic circuit $B_{\ell_0}$ of size at most $\ell_0^k$ that agrees with the language $L$ over $\{0,1\}^{\ell_0}$, i.e., $B_{\ell_0}(x) = M(x)$ for all $x \in \{0,1\}^{\ell_0}$.
\end{clm}

Note that the existence of $B_{\ell_0}$ is in contradiction with the existence of the refuter $R_{\ell_0}$, since there is no string $x_{\ell_0}$ such that $B_{\ell_0}(x_{\ell_0}) \neq M(x_{\ell_0})$. Consequently, in order to complete the proof of \Cref{thm:norefuter}, it is sufficient to establish \Cref{claim:upper_from_refuter}.

\begin{proof}[Proof of \Cref{claim:upper_from_refuter}] 
\noindent We will consider a large enough input length $\ell_0 \geq n_0$ specified later in the proof. First, we describe the pseudocode of a (recursively defined) deterministic circuit $B_{\ell_0}(x)$ that computes as follows:
%\footnote{We often write $M_u$ to emphasize that we run the machine $M$ on inputs of length $u$.}
\begin{enumerate}
\item $B_{\ell_0}$ is given a string $x_{\ell_0} \in \{0,1\}^{\ell_0}$ as input. 
    \item It computes the description of a co-nondeterministic circuit $E_{\ell_1}(z)$ operating on inputs of length $\ell_1$ that ignores its input string $z$ and satisfies $E_{\ell_1}(z) \eqdef 1-M(x_{\ell_0})$.
    
    (The value $\ell_1 < \ell_0$ will be defined below.)
    \item $B_{\ell_0}$ obtains $x_{\ell_1} \eqdef R_{\ell_1}(E_{\ell_1})$.
    
    (Note that if $\mathsf{size}(E_{\ell_1}) \leq \ell_1^k$ then $M(x_{\ell_1}) = 1 -  E_{\ell_1}(x_{\ell_1}) = 1 - (1-M(x_{\ell_0})) = M(x_{\ell_0})$.)
    \item Finally, it computes $b_{\ell_1} \eqdef M(x_{\ell_1})$, and uses this bit to decide $M(x_{\ell_0})$.

    (As explained below, the key point of the construction is that we can (deterministically) compute the bit $b_{\ell_1}$ in a recursive fashion via Steps 1--3 using the sequence of refuters.)
\end{enumerate}

\vspace{0.1cm}

Before elaborating on the recursion performed in Step 4, we discuss the complexity parameters involved. Let $t(\ell_1)$ denote the circuit complexity of computing the bit $b_{\ell_1} \eqdef M(x_{\ell_1})$ on an arbitrary input $x_{\ell_1}$, i.e., the complexity of deciding if $x_{\ell_1} \in L$ over strings of length $\ell_1$. As explained below, in order for $B_{\ell_0}$ to be correct and of the desired size, it is sufficient that:
\begin{itemize}
    \item $\mathsf{size}(E_{\ell_1}) \approx \ell_0 \leq \ell_1^k$ (i.e.,~the refuter receives a circuit of bounded size in Step 3), where we have used that $M$ runs in nondeterministic linear time and omitted polylog factors. 
    
    (In our analysis below, we can assume that the description of $E_{\ell_1}$ can be computed in size at most $\ell_0^k/4$, since $k > 1$ and $M$ runs in linear time.)
    \item We need that $\ell_1 \geq n_0$, so the output of $R_{\ell_1}(E_{\ell_1})$ is defined in Step $3$. 
    \item The circuit size of the refuter in Step $3$, which is upper bounded by $\ell_1^c$, satisfies $\ell_1^c \leq \ell_0^k/4$.
    \item The circuit size $t(\ell_1)$ needed to compute $b_{\ell_1}$ in Step 4 satisfies $t(\ell_1) \leq \ell_0^k/4$.
\end{itemize}
If these conditions are met, $B_{\ell_0}(x) = M(x)$ for all $x \in \{0,1\}^{\ell_0}$, and its overall size $t(\ell_0)$ is strictly less than $\ell_0^k$ (with some room to spare that will be handy later in the proof), since
$$
t(\ell_0) \leq \ell_0^k/4 + \ell_0^k/4 + t(\ell_1) \leq \ell_0^k/4 + \ell_0^k/4 + \ell_0^k/4 < \ell_0^k.
$$

\noindent \textbf{Constraints.} The conditions described above yield the following inequalities (omitting some low order terms):
$$
\ell_0^{1/k} \leq \ell_1 \leq \ell_0^{k/c}\;\;(\text{thus}~c < k^2) \quad \text{and} \quad t(\ell_1) \leq \ell_0^k/4.
$$
Recall that the condition $c < k^2$ is one of our assumptions. Jumping ahead, we will define $\ell_1$ as a function of $\ell_0$, $k$, and $c$, and employ a recursive approach to compute $b_{\ell_1}$ so that the conditions are satisfied.\\ 

\noindent \textbf{Key insight.} Note that in order to compute $b_{\ell_1} = M(x_{\ell_1})$ we must solve the \emph{same problem} on a \emph{smaller input length}, i.e., we would like to design a circuit $B_{\ell_1}$ that computes $L$ on input length $\ell_1 \ll  \ell_0$. Consequently, using that we have refuter circuits for all input lengths $\geq n_0$, it is enough to iterate the same construction!\\

\noindent \textbf{Recursion.} The circuit $B_{\ell_1}$ computes analogously to $B_{\ell_0}$, by considering a smaller input length $\ell_2 \ll \ell_1$ and a corresponding co-nondeterministic circuit $B_{\ell_2}$. Similarly to the analysis from above, we obtain the following constraints to guarantee its correctness and the desired circuit size bound (again omitting small order factors to focus on the relevant asymptotics):
$$
\mathsf{size}(E_{\ell_2}) \approx \ell_1 \leq \ell_2^k \quad \text{and} \quad \ell_2^c \leq \ell_1^k/4 \quad \text{and} \quad  t(\ell_2) \leq \ell_1^k/4.
$$
This forces that
$$
\ell_1^{1/k} \leq \ell_2 \leq \ell_1^{k/c} \quad \text{and} \quad t(\ell_1) \leq \ell_1^k/4 + \ell_1^k/4 + t(\ell_2) < 3 \cdot \ell_1^k/4 \leq \ell_0^k/4,
$$
where we assumed that $t(\ell_2) \leq \ell_1^k/4$ and used that $\ell_1 \ll \ell_0$. In other words, we need 
$$
\ell_1^{1/k} \leq \ell_2 \leq \ell_1^{k/c} \;\; (\text{thus}~c < k^2) \quad \text{and} \quad t(\ell_2) \leq \ell_1^k/4.
$$
 
\noindent \textbf{Parameters and base case.} Consider the input lengths $\ell_0 > \ell_1 > \ldots >  \ell_d$ explored in this way, together with the corresponding circuits $B_{\ell_0}, \ldots, B_{\ell_d}$, where each $B_{\ell_i}$ contains $B_{\ell_{i + 1}}$ as a subroutine. In other words, $B_{\ell_0}$ appears close to the input string, followed by $B_{\ell_1}$, and so on. (The string $x_{\ell_{i+1}}$ computed by $B_{\ell_i}$ serves as the input string to $B_{\ell_{i+1}}$.) We start with an input length $\ell_0$ sufficiently larger than $n_0$ so that all calls to the non-uniform refuter $R$ consider input lengths not smaller than $n_0$. In order to satisfy the inequalities from above, our parameters are defined as follows:\\

\noindent \emph{-- Sequence of input lengths.} We let $\ell_{i + 1} \eqdef \ell_i^{1/k}$. Therefore, $\ell_d = \ell_0^{1/k^d}$.\\

\noindent \emph{-- Number of stages.} We take $d$ large enough, so that $\ell_d = \ell_0^{1/k^d} = \log \ell_0$, i.e., $d \eqdef (\log \log \ell_0 - \log \log \ell_d)/\log k  = O(\log \log \ell_0)$.\\

\noindent \emph{-- Initial input length $\ell_0$.} We want $\ell_d = \log \ell_0 \geq n_0$, so we set $\ell_0 \eqdef 2^{n_0}$.\\

\noindent In the base case (input length $\ell_d$ and circuit $B_{\ell_d}$), we simply consult the hardcoded truthtable of the language $L$ computed by machine $M$ on inputs of length $\ell_d  = \log \ell_0$. The truthtable of $L$ on input length $\log \ell_0$ can be nonuniformly stored using $\ell_0$ bits. This can be seen as an overhead in the final size of $B_{\ell_0}$ that is of order $\ell_0 \leq \ell_0^k/4$, where this inequality uses that $k > 1$ and assumes that $\ell_0 \geq n_0$ is large enough. Since the overall size bound for $B_{\ell_0}$ is given by a simple additive  function obtained from the concatenation of a sequence of circuits, it is not hard to see that its total size 
is at most $\ell_0^k$, as desired.
\end{proof}
As explained above, this completes the proof of \Cref{thm:norefuter}.
\end{proof}

The next lemma will be used in the proof of \Cref{thm:intro_unprovIS12} Item (\emph{ii}).

\begin{lem}\label{lmm: bootstrapping refuter}
    Let $L\in\NTIME[n^c]$ for some constant $c \geq 1$. The following statements hold:
    \begin{enumerate}[label=(\roman*)]
        \item If there is a polynomial-time refuter $R$ for the lower bound $L\notin \io\coNSIZE[s(n)]$ for some monotone time-constructible function $s \colon \mathbb{N} \to \mathbb{N}$ such that $\omega(n^c \cdot \log n)\le s(n)\le \poly(n)$, then $L\in\P$.
        \item If there is a polynomial-size non-uniform refuter $R$ for the lower bound $L\notin \linebreak[5] \io\coNSIZE[s(n)]$ for some monotone time-constructible function $s \colon \mathbb{N} \to \mathbb{N}$ such that $\omega(n^c \cdot \log n)\le s(n)\le \poly(n)$, then $L\in\P/\poly$.
    \end{enumerate}
\end{lem}

\begin{proof} The argument is similar to the construction described above. In order to establish the first item, consider the following polynomial-time algorithm that aims to solve $L$. On a given input instance $x\in\zo^n$ for $L$, we construct a $\coNSIZE[O(n^c \cdot \log n)]$-circuit $D_x:\zo^{n/2} \times \{0,1\}^{n^c}\to\zo$ of the form $D_x(u, v)$ (where $v$ is the nondeterministic input string) that ignores its primary input $u \in \{0,1\}^{n/2}$ and computes according to $1-L(x)$. Since $L \in \mathsf{NTIME}[n^c]$, the transformation from machines to circuits guarantees that $D_x(u, \cdot)$ is a co-nondeterministic circuit of size at most $O(n^c \cdot \log n) \leq s(n/2)$, assuming that $n$ is sufficiently large. Moreover, a description of $D_x$ can be computed from $x$ and from the nondeterministic machine for $L$ in time polynomial in $n$.

Using the polynomial-time refuter $R$ on input $(1^{n/2},D_x)$ and the size upper bound for $D_x$, we can find a string $z_1\in\zo^{n/2}$ such that $D_x(z_1)\ne L(z_1)$. Using the definition of the circuit $D_x$, which as a co-nondeterministic circuit satisfies $D_x(u) = 1 - L(x)$ for every input $u \in \{0,1\}^{n/2}$, this implies that $$L(z_1)= 1 - D_x(z_1) = 1-(1 - L(x))=L(x)\;.$$ 

To sum up, in time polynomial in $n$, we have reduced the problem of deciding $L$ on $x \in \{0,1\}^n$ to that of deciding $L$ on $z_1 \in \{0,1\}^{n/2}$. We now  recursively evaluate $L(z_1)$ in a similar fashion until the input length is smaller than a large enough constant $C$. In other words, we produce a sequence $z_1, \ldots, z_k$ of inputs, where each $|z_i| = n/2^i$, $k \leq \log n$, $|z_k| \leq C$, and $$L(x) = L(z_1) = \ldots = L(z_k)\;.$$ Since $L(z_k)$ can be computed in constant time by brute force and there are at most $\log n$ stages of the recursion, it follows that we can decide $L(x)$ in time polynomial in $n = |x|$, i.e., $L \in \mathsf{P}$.  

The proof of the second item is completely analogous to the proof of the first item, i.e., it is sufficient to consider circuit size instead of running time.
\end{proof}

We obtain the following consequence.

\begin{thm}[Restatement of \Cref{thm:intro_refuters} Item (\emph{ii})]
There is an integer $k \geq 1$ such that there is no non-uniform refuter $R(1^n,E_n)$ for $\mathsf{SAT}_n$  against $E_n \in \mathsf{coNSIZE}[n^k]$ of polynomial circuit size which succeeds on every large enough input length.
\end{thm}

\begin{proof}
    Assume such a polynomial size refuter exists for $k = 2$, since we are done otherwise. Using that $\mathsf{SAT} \in \mathsf{NTIME}[n \cdot \mathsf{poly}(\log n)]$ and setting $s(n) = n^2$, it follows from \Cref{lmm: bootstrapping refuter} that $\mathsf{SAT} \in \mathsf{P}/\poly$. As a consequence, there is a constant $c \in \mathbb{N}$ and a sequence of non-uniform (deterministic) circuits of size $n^c$ that compute $\mathsf{SAT}_n$ on every large enough input length $n$. But then there is no refuter witnessing that $\mathsf{SAT}_n \notin \mathsf{coNSIZE}[n^k]$ when $k = c + 1$, since this  lower bound is simply false when the input length is large enough.
\end{proof}

\subsection{Unprovability of lower bounds via refuter lower bounds}

In this section, we prove each item of \Cref{thm:intro_unprovIS12}.  We will need the following witnessing result for the intuitionistic theory $\ISonetwo$.

\begin{thm}[Witnessing Theorem for $\ISonetwo$ \citep{DBLP:conf/coco/Buss86a, MR1241248}] \label{thm:ISwitnessing} Let $\varphi$ be an arbitrary formula, and suppose that $$\mathsf{IS}^1_2 \vdash \forall x \;\exists y~\varphi(x,y)\,.$$ Then there is a polynomial-time computable function $f$ such that $$\mathbb{N} \models \forall a~\varphi(a,f(a))\,.$$
Furthermore, there is a $\Sigma^{b+}_1$-formula $\psi(x,y)$ such that:
\begin{enumerate}[label=(\roman*)]
    \item $\mathsf{IS}^1_2 \vdash \forall x\,\forall y\,(\psi(x,y) \rightarrow \varphi(x,y))\,$.
    \item $\mathsf{IS}^1_2 \vdash \forall x\,\forall y\,\forall z\,(\psi(x,y) \wedge \psi(x,z) \rightarrow y = z)\,$.
    \item $\mathsf{IS}^1_2 \vdash \forall x\,\exists y\,\psi(x,y)\,$.
\end{enumerate}
\end{thm}

In particular, \Cref{thm:ISwitnessing} shows that the outermost existential quantifier of a sentence of arbitrary quantifier complexity provable in $\mathsf{IS}^1_2$ can be efficiently witnessed by a polynomial-time function.

\begin{thm}[Restatement of \Cref{thm:intro_unprovIS12} Part (\emph{i})] \label{thm:unprovIS12}
    Let $\delta > 0$ be a rational number, $n_0 \in \mathbb{N}$, and $M$ be a polynomial-time nondetermistic  machine. Then
    $
  \mathsf{IS}^1_2 \nvdash   \mathsf{LB}^{\mathsf{exp}}(M,2^{n^{\delta}},n_0)\;.
$
\end{thm}

\begin{proof} We argue by contradiction. Under the assumption that 
    $$
\mathsf{IS}^1_2 \vdash   \mathsf{LB}^{\mathsf{exp}}(M,2^{n^{\delta}},n_0)
    $$
    for a choice of $M$, $\delta > 0$, and $n_0$, it follows from \Cref{thm:ISwitnessing} that there is a refuter $R(1^n, D)$ that runs in time $2^{O(n)}$ and outputs an input $x \in \{0,1\}^n$ such that $\mathsf{Error}_M(D,n,x)$, whenever $D$ is a co-nondeterministic circuit of size at most $2^{n^{\delta}}$ and $n \geq n_0$. By \Cref{thm:UB_from_refuter}, for any choice of $\varepsilon > 0$, we get that $L(M) \in \mathsf{DTIME}[2^{n^{\varepsilon}}]$. Taking $\varepsilon < \delta$, this upper bound and the provability of $\mathsf{LB}^{\mathsf{exp}}(M,2^{n^{\delta}},n_0)$ contradict the soundness of $\mathrm{IS}^1_2$.
\end{proof}

\begin{thm}[Restatement of \Cref{thm:intro_unprovIS12} Part (\emph{ii})] \label{thm:unprovIS12poly}
   Let $n_0 \in \mathbb{N}$ and $M$ be a polynomial-time nondetermistic  machine. Then there is an integer $k  \geq 1$ such that
    $
  \mathsf{IS}^1_2 \nvdash   \mathsf{LB}^{\mathsf{poly}}(M,n^k,n_0)\;.
  $
\end{thm}

\begin{proof} Let $n^c$ be an upper bound on the nondeterministic time complexity of $M$. Towards a contradiction, assume that $\mathsf{IS}^1_2 \vdash   \mathsf{LB}^{\mathsf{poly}}(M,n^k,n_0)$ for every $k \geq 1$. In particular, this holds for $k = 2c$. It follows from \Cref{thm:ISwitnessing} that there is a refuter $R(1^n, D)$ that runs in time $\mathsf{poly}(n)$ and outputs an input $x \in \{0,1\}^n$ such that $\mathsf{Error}_M(D,n,x)$, whenever $D$ is a co-nondeterministic circuit of size at most $n^k$ and $n \geq n_0$. Consequently, by \Cref{lmm: bootstrapping refuter}, we get that $L(M) \in \P$. In particular, the sentence $\mathsf{LB}^{\mathsf{poly}}(M,n^k,n_0)$ is \emph{false} for some large enough constant $k$. This and the assumption that $\mathsf{IS}^1_2 \vdash   \mathsf{LB}^{\mathsf{poly}}(M,n^k,n_0)$ for every $k \geq 1$  contradict the soundness of $\IS^1_2$. 
\end{proof}

\section{Unprovability of Upper Bounds in \texorpdfstring{$\mathsf{IS}^1_2$}{IS12}} \label{ref:unprov_upper_bounds}

In this section, we prove unconditionally that a natural formalization of the circuit \emph{upper bound} $\NP\subseteq\coNSIZE[n^k]$ is unprovable in $\IS^1_2$, for every fixed constant $k$. 
\newcommand{\UB}{\mathsf{UB}}

\subsection{Unconditional uniform lower bound against co-nondeterministic circuits}

\newcommand{\PUniform}{\mathsf{P}\text{-}\mathsf{uniform}}

We first prove that $\NP\nsubseteq\PUniform~\coNSIZE[n^k]$, for every constant $k$. That is, for each $k$, there is a language $L\in\NP$ that cannot be decided by any polynomial-time uniform family of co-nondeterministic circuits of size $n^k$. This uniform lower bound is obtained through an adaptation of a technique of Santhanam and Williams \cite{DBLP:journals/cc/SanthanamW14} showing that $\P\nsubseteq\PUniform~\SIZE[n^k]$. 

\newclass{\coNTIME}{coNTIME}

First, we will need the following standard ``no complementary speedup theorem''. Following standard notation from complexity theory, we use $\mathsf{NTIME}[t(n)]/a(n)$ to denote the set of languages that can de decided in nondeterministic time $t(n)$ using $a(n)$ advice bits per input length.

\begin{thm}[Folklore]\label{thm: no compl speedup}
    For every constant $b\ge 1$, $\coNTIME[n^{b+1}]\nsubseteq\NTIME[n^b]/o(n)$.
\end{thm}

\begin{proof}[Proof Sketch]
    Fix any constant $b\ge 1$. Consider the following co-nondeterministic Turing machine $M$ that runs in time $O(n^{b+1})$: On any input of the form $x=(\hat M,\alpha,\pi)\in\zo^n$, where $\hat M$ is interpreted as the encoding of a (clocked) nondeterministic Turing machine running in time $O(n^b)$, $\alpha_n$ is interpreted as the advice to $\hat M$ on input length $n$, and $\pi\in\{0\}^*$ is a padding string, $M$ simulates $\hat M(x)_{\alpha_n}$ (i.e., $\hat M$ on input $x$ with advice string $\alpha_n$) and accepts if and only if $\hat M(x)_{\alpha_n}$ rejects. It is not hard to show that the language $L(M)\notin\NTIME[n^b]/o(n)$. 
\end{proof}

Next, we introduce an auxiliary definition and establish a proposition that will be useful. 

\newfunc{\pDCL}{pDCL}
\newfunc{\nDCL}{nDCL}
\newcommand{\de}{\text{-}}

\begin{defi}
    Let $C=\{C_n\}_{n\ge 1}$ be a family of polynomial-size nondeterministic circuits, and $m=m(n)$ be a constructive function. The positive and negative $m$-padded \emph{direct connect language} for $C$, short for $m\de\pDCL(C)$ and $m\de\nDCL(C)$, are defined as follows: 
    \begin{align*}
        & m\de\pDCL(C)\eqdef \{(n,1^m,i)\mid \langle C_n\rangle_i=1\}, \\ 
        & m\de\nDCL(C)\eqdef \{(n,1^m,i)\mid \langle C_n\rangle_i=0\},
    \end{align*}
    where $\langle C_n\rangle$ is the string that encodes the circuit $C_n$. 
\end{defi}

\begin{prop}\label{prop: dcl in p}
    If $C=\{C_n\}_{n\ge 1}$ is a $\PUniform$ family of \emph{(}nondeterministic\emph{)} circuits, the padded direct connect languages $n^\eps\de\pDCL(C),n^\eps\de\nDCL(C)\in\P$ for any fixed constant $\eps\in(0,1)$.
\end{prop}

\begin{proof}
    Given $(n,1^{n^\eps},i)$, one can first print the description $y=\langle C_n\rangle$ of $C_n$ in $\poly(n)$ time, then check whether the $i$-th bit is $0$ or $1$. Since the input length is at least $n^\eps$, this algorithm runs in polynomial time. 
\end{proof}

\begin{thm}\label{thm: uniform-lb}
For every positive integer $k$,     $\NP\nsubseteq\PUniform~\coNSIZE[n^k]$.
\end{thm}

\begin{proof}
    We argue that $\coNP\nsubseteq\PUniform~\NSIZE[n^k]$. The theorem   follows from this separation by a simple complementation argument. Let $k$ be any constant, and $H_b$ be the hard language in \Cref{thm: no compl speedup} for some $b>k$ to be determined later. Towards a contradiction, we assume that for every language $L\in\coNP$, there is a family of $\PUniform$ nondeterministic circuits $C=\{C_n\}_{n\ge 1}$ of size $c n^k$ that decides $L$, where $c$ is an arbitrary constant that can depend on $L$. In particular, this uniform fixed polynomial upper bound holds for $H_b$. We will use the fixed polynomial upper bound for $\coNP$ to optimize $H_b$ to the extent that it violates \Cref{thm: no compl speedup}. 
    \begin{enumerate}
        \item (\emph{Pad Down}). Let $\eps\eqdef 1/2k$ and $C=\{C_n\}_{n\ge 1}$ be the $\PUniform$ family of nondeterministic circuits of size $c n^k$ that decides $H_b$. By \Cref{prop: dcl in p}, we know that $n^\eps\de\pDCL(C)$ and $n^\eps\de\nDCL(C)$ are in $\P$, and therefore also in $\coNP$. 
        \item (\emph{Compress} $C_n$). Let $m=m(n)\le O(\log n)+n^\eps$ be the length of the input $(n,1^{n^\eps},i)$ for $n^\eps\de\pDCL(C)$ and $n^\eps\de\nDCL(C)$. Since $n^\eps\de\pDCL(C)\in\coNP$, by the upper bound for $\coNP$, we get that there is a family of nondeterministic circuits $D^p=\{D^p_m\}_{m\ge 1}$ of size $c' m^k$ that decides $n^\eps\de\pDCL(C)$. Similarly, there is a family of nondeterministic circuits $D^n=\{D^n_m\}_{m\ge 1}$ of size $c'' m^k$ that decides $n^\eps\de\nDCL(C)$. Note that the sizes of both circuits $D^p_m$ and $D^n_m$ are $O(m^k) \le O(n^{\eps k})=O(\sqrt n)$, which gives a succinct representation of $C_n$. 
        \item (\emph{Speedup with Advice}). Now we speedup the computation of $H_b$ with $C_n$, $D^p_m$, and $D^n_m$. Let $M'_{\alpha_n}$ be the following nondeterministic algorithm with advice $\{\alpha_n\}_{n\ge 1}$ of length $o(n)$:
        \begin{itemize}
            \item The advice is defined as $\alpha_n\eqdef (D_m^p,D_m^n)$.
            \item Let $\alpha_n=(D_m^p,D^m_n)$ be the advice and $x\in\zo^n$ be the input. Let $\ell\eqdef |\langle C_n\rangle|$. Note that since $C_n$ is of size $O(n^k)$, $\ell \le \tilde{O}(n^{k})$. The algorithm nondeterministically guesses a string $y\in\zo^{\ell}$ for every $i\in[\ell]$, then verifies that for every $i\in[\ell]$, $y_i=\langle C_n\rangle_i$. Concretely, for every $i\in[\ell]$, the algorithm works as follows: if $y_i=1$, it simulates $D_m^p(n,1^{n^\eps},i)$ and immediately rejects if $D_m^p$ rejects; otherwise, the algorithm simulates $D_m^n(n,1^{n^\eps},i)$ and immediately rejects if $D_m^n$ rejects. Note that since both $D_m^p$ and $D_m^n$ are nondeterministic circuits of size $O(\sqrt n)$, the running time for this step is $O(\ell)+\tilde O(\sqrt n) = \tilde O(n^k)$. After this step, we know that $y=\langle C_n\rangle$.  
            \item Finally, the algorithm simulates $C_n(x)$, which takes $\tilde O(n^k)$ time. 
        \end{itemize}  

        In summary, this algorithm takes an advice of length $o(n)$ and simulates $C_n(x)$ in time $\tilde O(n^k)$. Since $C_n$ decides $H_b$, this algorithm also decides $H_b$. Therefore, $H_b\in\NTIME[n^{k+1}]/o(n)$.  
    \end{enumerate} 
    This leads to a contradiction by choosing $b=k+2$. 
\end{proof}

\subsection{Unprovability of upper bounds via uniform lower bounds}

We now combine the uniform lower bound (\Cref{thm: uniform-lb}) and the witnessing theorem for $\mathsf{IS}^1_2$ (\Cref{thm:ISwitnessing}) to establish the unprovability of $\NP\subseteq\coNSIZE[n^k]$ in $\IS^1_2$. First, we explain how to formalize this upper bound statement. This is similar to a formalization in \citep{KO17}.

\paragraph{Formalization.} The fixed polynomial circuit upper bound $\NP\subseteq \coNSIZE[n^k]$ states that for every polynomial-time nondeterministic Turing machine $M$ and every input length $n$, there is a co-nondeterministic circuit $C$ of size $O(n^k)$ such that $C(x)=M(x)$. For each $k$, we capture the upper bound statement using a collection of sentences 
\[ 
    ``\NP\subseteq\coNSIZE[n^k]"\eqdef \{ \UB_M(k,c) \mid  M\text{ is an }\NP\text{ machine and}~c \in \mathbb{N}\},
\] 
where $\UB_M(k,c)$ is the sentence:
\begin{align*}
\UB_M(k,c)\eqdef \forall n\in\Log~\exists C\in\coNSIZE[cn^k]~\forall x\in\zo^n~\lnot\mathsf{Error}_M(C,n,x).
\end{align*}
The sentence $\UB_M(k,c)$ can be expressed in a natural way, similarly to the lower bound sentence $\mathsf{LB}^{\mathsf{poly}}$  described in \Cref{sec:prelim_intuitionistic}.

For a fixed $k$, we say that a theory $T$ proves $``\NP\subseteq \coNSIZE[n^k]"$ if for every polynomial-time nondeterministic Turing machine $M$ there is a constant $c$ such that $T\vdash\UB_M(k,c)$.\\

We are now ready to prove the main result of this section.

\begin{thm}\label{thm:unprovIS12upper}
    For every $k\in\bbN$, $\IS^1_2\nvdash``\NP\subseteq\coNSIZE[n^k]"$.
\end{thm}

\begin{proof}
    Towards a contradiction, we assume that $\IS^1_2\vdash``\NP\subseteq\coNSIZE[n^k]"$ for some $k\in\bbN$, that is, for every $\NP$ machine $M$ there is a constant $c_M$ such that $\IS^1_2\vdash \UB_M(k,c_M)$. For convenience, we write 
    \[
    \UB_M(k,c)=\forall n\in\Log~\exists C\in\coNSIZE[c_Mn^k]~\varphi(n,C),    
    \]
    where $\varphi(n,C)\eqdef \forall x\in\zo^n~\lnot\mathsf{Error}_M(C,n,x)$. By the witnessing theorem for $\mathsf{IS}^1_2$ (\Cref{thm:ISwitnessing}), for each $\NP$ machine $M$, there is a polynomial-time algorithm $A_M$ that takes $1^n$ as input and outputs a co-nondeterministic circuit $C_n$ of size at most $c_M n^k$ such that $\varphi(n,C_n)$ holds in the standard model $\bbN$. In other words, $C = \{C_n\}_{n \geq 1}$ is a $\mathsf{P}$-uniform co-nondeterministic circuit family that decides $L(M)$. (Note that the algorithm takes $1^n$ instead of $n$ as input as $\forall n\in\Log$ is a shorthand for $\forall v~\forall n=|v|$.) This immediately implies that every language in $\NP$ can be computable by a $\P$-uniform family of co-nondeterministic circuits of size $O(n^k)$, which is impossible by \Cref{thm: uniform-lb}.
\end{proof}

\section{Interpretation and Generalization of Our Results}\label{sec:unprov_upper_bound_extension}

We make two remarks about the interpretation and generalization of our results. 

\paragraph{Interpretation of the unprovability results.} In this paper, we established the unprovability of both co-nondeterministic circuit size upper bounds and lower bounds in the intuitionistic theory $\IS^1_2$. Roughly speaking, the standard interpretation of the results is that the question of whether $\NP$ requires large co-nondeterministic circuits cannot be  constructively resolved, either because of the \emph{lack of relevant non-logical axioms} or because of \emph{the absence of the law of excluded middle}. 

Due to the nature of intuitionistic logic, it could still be the case that while $\NP\subseteq\coNSIZE[n^k]$ is unprovable in $\IS^1_2$, $\lnot\lnot``\NP\subseteq\coNSIZE[n^k]"$ is indeed provable in $\IS^1_2$, as $\lnot\lnot\varphi\vdash \varphi$ is not an admissible inference rule in intuitionistic logic. As explained in more detail next, our results can be considered necessary steps towards the classical independence of worst-case lower bounds. (See also \citep[Section 6.2]{buss1990model} for related considerations in a different context.)

\paragraph{Robustness of our unprovability results.} Recall that the proofs of our unprovability results in \Cref{thm:unprovIS12}, \Cref{thm:unprovIS12poly}, and \Cref{thm:unprovIS12upper} only rely on the soundness of $\IS^1_2$ over the standard model and on the witnessing theorem to extract a polynomial-time algorithm for the \emph{outermost existential quantifier}. Therefore, the unprovability result holds even if one substitutes a subformula of the lower bound or upper bound sentences inside the outermost existential quantifier by any formula that coincides with the intended meaning over the standard model $\mathbb{N}$. 

In particular, the unprovability result holds even if any subformula inside the outermost existential quantifier is replaced by any classically equivalent formula. This suggests that our unprovability result holds in ``semi-classical'' setting, i.e., classical reasoning is allowed inside the outermost existential quantifier. 

\newcommand{\NN}{\mathsf{N}}

A more formal way to phrase the observation is via the well-known \emph{double-negation translation} (i.e.~G\"{o}del–Gentzen translation). The double-negation translation of a first-order sentence $\varphi$, denoted by $\varphi^\NN$, is defined inductively by the following rules: 
    \begin{itemize}
        \item $A^\NN\eqdef \lnot\lnot A$~for atomic $A$ (i.e.~$A$ is a predicate);
        \item $(\varphi\land\psi)^\NN\eqdef \varphi^\NN\land\psi^\NN$; $(\lnot\varphi)^\NN\eqdef \lnot\varphi^\NN$; $(\varphi\to\psi)^\NN\eqdef \varphi^\NN\to\psi^\NN$; 
        \item $(\varphi\lor\psi)^\NN\eqdef \lnot\lnot(\varphi^\NN\lor\psi^\NN)$; 
        \item $(\forall x~\varphi)^\NN\eqdef \forall x~\varphi^\NN$; 
        \item $(\exists x~\varphi)^\NN\eqdef \lnot\lnot\exists x~\varphi^\NN$. 
    \end{itemize}
For a set $\Pi$ of formulas, we define $\Pi^\NN\eqdef\{\varphi^\NN\mid \varphi\in\Pi\}$. 

\begin{thm}[see, e.g., \cite{buss1998handbook}]
For a set $\Pi$ of formulas and a formula $\varphi$, $\Pi$ proves $\varphi$ classically if and only if $\Pi^\NN$ proves $\varphi^\NN$ intuitionistically. In particular, $\varphi$ and $\psi$ are classically equivalent if and only if $\varphi^\NN$ and $\psi^\NN$ are intuitionistically equivalent. 
\end{thm}

\begin{cor}
    In \Cref{thm:unprovIS12}, \Cref{thm:unprovIS12poly}, and \Cref{thm:unprovIS12upper}, the corresponding unprovability result holds even if the subformula inside the outermost existential quantifier is replaced by its double-negation translation. 
\end{cor}

\ifthenelse{\value{anonymous}=0}{
\section*{Acknowledgement}
We would like to thank Erfan Khaniki and Dimitrios Tsintsilidas for discussions on proof complexity lower bounds in intuitionistic bounded arithmetic. We are also grateful to the anonymous reviewers for several comments that improved our presentation. Igor C.~Oliveira received support from the Royal Society University Research Fellowship URF$\setminus$R1$\setminus$191059; the UKRI Frontier Research Guarantee Grant EP/Y007999/1; and the Centre for Discrete Mathematics and its Applications (DIMAP) at the University of Warwick. Lijie Chen is supported by a Miller Research Fellowship. Jiatu Li is supported by MIT Akamai Presidential Fellowship. 
}{}

\bibliographystyle{alphaurl}
\bibliography{references.bib}

\appendix

\section{On the Unprovability of Lower Bounds Against Deterministic Circuits}\label{sec:lb_det}

This section provides an overview of a result from \citep{ghasemloo2013note} showing that, under an appropriate formalization, the unprovability in $\mathsf{IS}^1_2$ of worst-case lower bounds against \emph{deterministic} Boolean circuits  yields the unprovability of the same lower bound in $\mathsf{S}^1_2$. The necessary background on bounded arithmetic can be found in \Cref{sec:prelim}. In particular, the vocabulary of $\mathsf{S}^1_2$ and $\mathsf{IS}^1_2$ is discussed in \Cref{sec:prelim_intuitionistic}.

\paragraph*{Formalization.} For concreteness, we consider a constant $\delta > 0$ and focus on the size bound $2^{n^{\delta}}$. (The actual size bound is not relevant for the result, provided that it can be captured by a sharply bounded formula in the sense described below.)   We use $\mathsf{SAT} = \{\mathsf{SAT}_n\}$ to denote the decision version of the satisfiability problem for de Morgan Boolean formulas represented by $n$-bit strings. We consider a sentence $\mathsf{SAT}\text{-}\mathsf{DLB}^{\mathsf{exp}}(2^{n^{\varepsilon}}, n_0)$ which encodes that, for every input length $n \geq n_0$ and for every deterministic circuit $D$ of size at most $2^{n^{\delta}}$, there is an input $y$ such that $\mathsf{SAT}_n(y) \neq D(y)$. The formalization will require the following auxiliary formulas. (The variable $f$ appearing in some of them but not referred to is used to guarantee bounded quantification, as discussed later in this section.)\\ 

\noindent $\mathsf{ONE}(f,y)$. For variables $f$ (a truth-table with $|f| = 2^n$) and $y$ (an index/input of $f$), we use the formula $\mathsf{ONE}(f,y)$ to determine if the $y$-th entry of $f$ is $1$.\\

\noindent $\mathsf{SAT}(f,y,z)$. For variables $y$ (encoding a Boolean formula) and $z$ (encoding an assignment), the formula $\mathsf{SAT}(f,y,z)$ denotes that $y$ is satisfied by $z$.\\

\noindent $\mathsf{DECIDES}\text{-}\mathsf{SAT}(f,y)$. For variables $f$ and $y$, the formula $\mathsf{DECIDES}\text{-}\mathsf{SAT}(f,y)$ denotes \linebreak[5] $\mathsf{ONE}(f,y) \leftrightarrow \exists z \leq |f|\;\mathsf{SAT}(y,z)$, i.e., $f$ correctly decides the satisfiability of the formula encoded by $y$. (The intended interpretation is that $z$ is an integer of magnitude at most $2^n$ and consequently can be viewed as an $n$-bit string.)\\

\noindent $\mathsf{SIZE}(f,D,n)$. For a variable $D$ (encoding a deterministic Boolean circuit), $\mathsf{SIZE}(f,D,n)$ denotes the formula that checks if $D$ has $n$ inputs and at most $2^{n^{\delta}}$ gates.\\

\noindent $\mathsf{ERROR}(f,D,w,y)$. For variables $f$, $D$, $w$ (a transcript of $D$'s computation), and $y$,  $\mathsf{Error}(f,D,w,y)$ denotes the formula stating that $w$ correctly encodes the computation of $D$ on $y$ and $f(y) \neq D(y)$.\\

\noindent $\mathsf{INDEX}(f, W, y, w)$. For variables $W$ (a sequence of transcripts), $y$, and $w$, $\mathsf{INDEX}(f, W, y, w)$ denotes that the $y$-th element of $W$ is $w$.\\

\noindent Some of these formulas might require the specification of additional sub-formulas in order to  capture their intended behavior. Most importantly, we note that all these formulas admit \emph{sharply bounded} descriptions due to the presence of the variable $f$ (we will use $n = ||f||)$ and the explicitly provided candidate transcript $w$ (whose correctness is easy to check). (This claim can be somewhat tedious to check; for additional details, see a similar presentation in \citep{ghasemloo2013note}.) Next, we let $\mathsf{SAT}\text{-}\mathsf{DLB}^{\mathsf{exp}}(2^{n^{\delta}}, n_0)$ denote the following sentence:
\begin{eqnarray} & \nonumber
    \forall f\;\forall n\;\forall D\;\forall W\;\forall w\;\exists y \leq |f|\;\\ \nonumber
   &  \Big (n = ||f|| \wedge n \geq n_0 \wedge \mathsf{DECIDES}\text{-}\mathsf{SAT}(f,y) \wedge \mathsf{SIZE}(f,D,n) \wedge \mathsf{INDEX}(f,W,y,w) \Big )\\ & \nonumber
    \rightarrow\\ \nonumber
   & \mathsf{ERROR}(f,D,w,y)\;.
\end{eqnarray}
Observe that $\mathsf{SAT}\text{-}\mathsf{DLB}^{\mathsf{exp}}(2^{n^{\delta}}, n_0)$ is the universal closure of a sharply bounded formula, which is in particular a $\forall \Sigma^b_1$-formula. It is not hard to see that it correctly captures (over the standard model $\mathbb{N}$) the statement that $\mathsf{SAT}_n \notin \mathsf{SIZE}[2^{n^{\delta}}]$ for all $n \geq n_0$.\\

Finally, we will make use of the following conservation result.

\begin{thm}[Avigad {\citep[Theorem 3.17]{DBLP:journals/jsyml/Avigad00}}] 
    $\mathsf{S}^1_2$ is conservative over $\mathsf{IS}^1_2$ for $\forall \Sigma^b_1$ sentences.
\end{thm}

As a consequence of this result and of the quantifier complexity of the formalization, the provability of $\mathsf{SAT}\text{-}\mathsf{DLB}^{\mathsf{exp}}(2^{n^{\delta}}, n_0)$ in $\Sonetwo$ yields its provability in $\ISonetwo$. In other words, if the lower bound sentence in unprovable in $\ISonetwo$, it is also  unprovable in $\Sonetwo$:

\begin{thm}\label{thm:conservation}
    For every rational $\delta > 0$ and $n_0 \in \mathbb{N}$, if $\mathsf{IS}^1_2 \nvdash \mathsf{SAT}\text{-}\mathsf{DLB}^{\mathsf{exp}}(2^{n^{\delta}}, n_0)$ then $\mathsf{S}^1_2 \nvdash \mathsf{SAT}\text{-}\mathsf{DLB}^{\mathsf{exp}}(2^{n^{\delta}}, n_0)$.
\end{thm}

This result should be contrasted with \Cref{thm:intro_unprovIS12} Item (\emph{i}), which establishes in particular the unprovability in $\mathsf{IS}^1_2$ of a \emph{co-nondeterministic} size lower bound for $\mathsf{SAT}$. Note that the quantifier complexity of the sentences $\mathsf{LB}^{\mathsf{exp}}$ and $\mathsf{LB}^{\mathsf{poly}}$ from \Cref{thm:intro_unprovIS12} does not allow us to invoke \Cref{thm:conservation} (nor an extension of this conservation result from \citep{coquand1999new}) to derive an unprovability result for $\mathsf{S}^1_2$.

%\newpage

%\input{diagram}

\end{document}